\documentclass[aps,prc, groupedaddress,superscriptaddress,twocolumn,showpacs]{revtex4}
\usepackage{graphicx}
\usepackage{longtable}
\usepackage{array}

\begin{document}

\author {Yu.~M.~Gavriljuk}
\author {V~.N.~Gavrin}
\author {A.~M.~Gangapshev}
\author {V.~V.~Kazalov}
\author {V.~V.~Kuzminov}
\affiliation{Baksan Neutrino Observatory INR RAS,  Russia}

\author {S.~I.~Panasenko}
\author {S.~S.~Ratkevich}
\affiliation{Karazin Kharkiv National University, Ukraine}

\date{\today}

\begin{abstract}
The construction of an ion pulse ionization chamber aimed at
measuring ultra-low levels of surface alpha-activity of
different samples is described. The results of measurement
carried out with alpha-source and copper samples and
light-reflecting film VM2000 are presented.
\end{abstract}


\title{\bf
Measurement of surface alpha-acrivity of different \\ samples with
ion pulse ionization chamber}

\maketitle

\section{\label{Intr}Introduction}

In construction of up-to-date low background detectors
searching for WIMPs and double beta-decay of different isotopes
it is necessary to have materials with the lowest possible
content of natural radioactive elements $^{238}$U, $^{232}$Th,
their daughter decay products and  $^{40}$K. Ultra-low
background semiconductor detectors are usually used to measure
radioactive isotopes in the selected material by registering
gamma radiation produced in the beta-decay process. The
concentration of the decaying isotopes is calculated by taking
into account the efficiency of registration of gamma radiation
for the given energy. In the decay chains of $^{238}$U,
$^{232}$Th, it is not all isotopes that undergo beta-decay with
gamma quanta emission. Some part of the isotopes decays with
alpha-particles emission. Calculation of the latter isotopes
content using gamma radiation of their parent nuclei is based
on the assumption of secular equilibrium in the chain. However,
this condition may not be always satisfied. For example, it is
known [1] that in the process of pure metal production from
crude ore a significant violation of equilibrium occurs in the
rows of $^{238}$U, $^{232}$Th due to the predominant
elimination of isotopes $^{226}$Ra (T$_{1/2}=1600$ yrs) and
$^{228}$Ra (T$_{1/2}=5.7$ yrs), respectively. In the finished
metal the radium content could be 1000 times less than the
equilibrium one [2]. The basic isotopes which emit gamma rays
are the products of radium daughters decays. Therefore, in case
of using their gamma emission in determination of $^{238}$U and
$^{232}$Th content, under the assumption of equilibrium, the
evaluated value would be significantly lower than the real one.
Moreover, if the material under consideration has been kept
long in the environment with high content of $^{222}$Rn, there
could be accumulated the superfluous amount of $^{210}$Pb atoms
(T$_{1/2}$=22.3 yrs) in comparison with the equilibrium
content. $^{210}$Pb and daughter products decay would give
additional background of electrons and alpha-particles.
Therefore, it is necessary to measure simultaneously surface
alpha-activity and gamma ray background for the material under
study. Ion pulse ionization chamber (IPIC) with ion collection
could be used for such a purpose.

A gas detector of such a type collects all the charge generated
in the working gas by the ionized particle for any type and
mobility of a carrier. This characteristic distinguishes IPIC
from other detectors with electronic collection where one needs
to use ultra-pure gases having no electronegative admixtures
capable of capturing electrons. In case of electron collection
detectors, to obtain good energy resolution one needs to insert
screen grids separating drift and register gaps. A grid allows
one to eliminate dependence of the signal's amplitude, induced
by the moving in the electric field ionization electrons, of
the distance between the ionizing track and the collector.
There is no necessity to insert such grids in case of IPIC and
there are no special requirements for the materials used in the
fiducial volume of IPIC as to the level and composition of the
gases emitted except radon itself.

A possibility to obtain  good enough energy resolution for
alpha-decays of $^{222}$Rn and its daughter products registered
in IPIC filled with air is presented in work [3].

\section{\label{sec2} Chamber's design and construction}

\begin{figure}
\includegraphics[width=\linewidth]{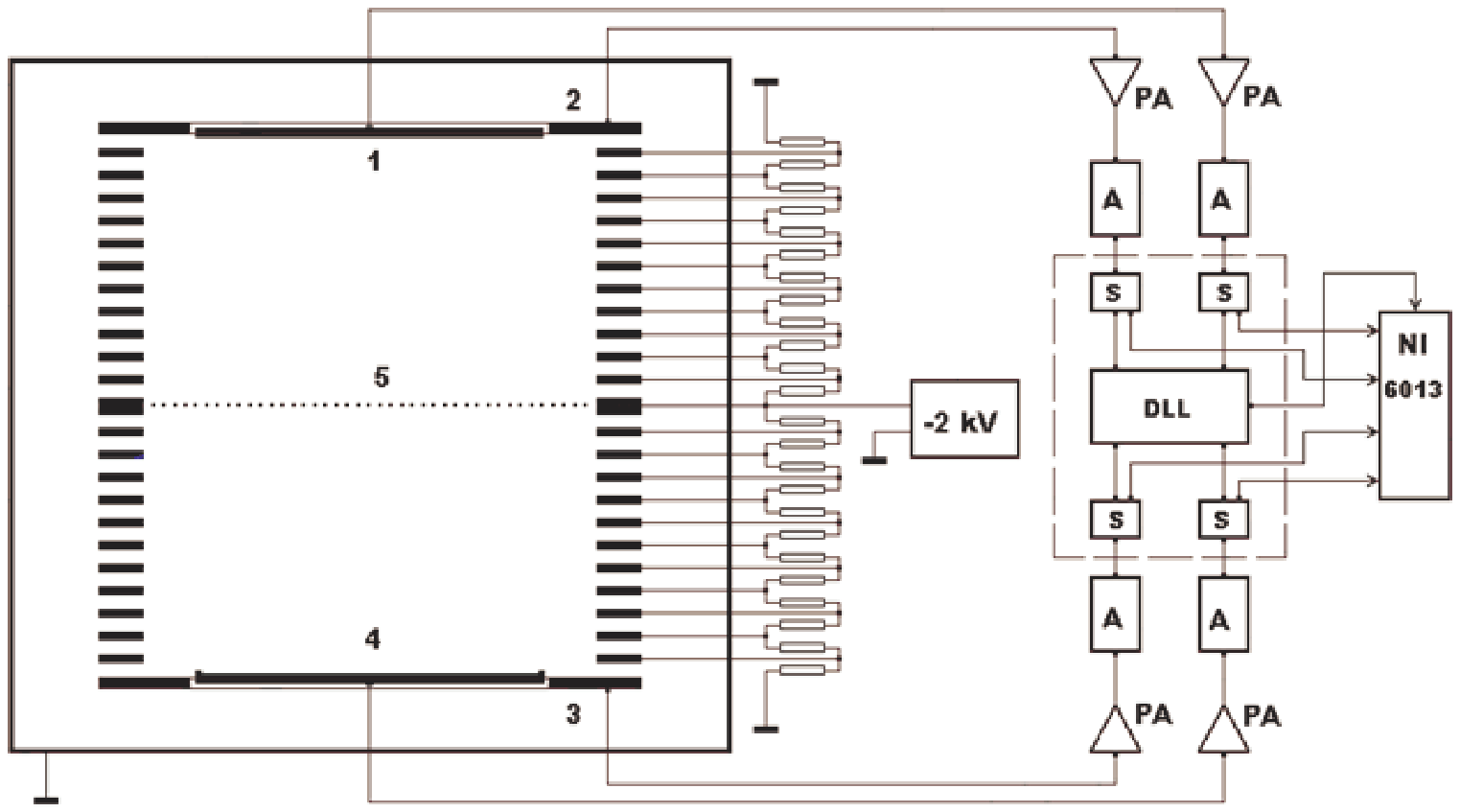}
 \caption{Schematic
cross-sectional view and connection of IPIC to the high-voltage
resistance divider 1 - upper central anode; 2 - upper circular
anode; 3 - lower circular anode; 4 - lower central anode;  5 -
cathode.}
\label{pic1}
\end{figure}

The diagram of the cross-section of IPIC and its circuit are
presented in fig.\ref{pic1}. Chamber consists of two identical
cylindrical coaxial sections separated by a high-voltage grid
cathode electrode (5). The grid is made of parallel copper
wires of 0.08 mm diameter with 2 mm spacing. High voltage of -2
kV is applied to it. The uniform drift field in each section is
produced by a set of forming copper circular electrodes (6),
connected to the high-voltage resistance divider. Collectors
(anodes) of the upper and lower sections are electrically
divided to form the inner and outer parts, disk-shaped and
circular, respectively.
In fig.\ref{pic1} they are denoted as (1), (2), and (4), (3),
respectively. Electrode (4) has a cylindrical cavity to
accommodate the sample under study. In case when sample is made
of non-conductor material it is covered with a copper grid
(diameter of the wire 0.08 mm, spacing 8õ8 mm), contacting the
skirting of the electrode. Total diameter of the drift spacing
is 126 mm, its height is 74 mm; the diameter of the electrode
is 91 mm.  The diameter and digging depth of the cavity for a
sample is 84 mm and 3 mm, respectively. Geometrical area of the
sample is 55.4 cm$^2$. The height for a drift spacing for the
given potential on the cathode and gas pressure determines full
time charge collection and consequently the amplitude and
duration of the current pulse. To improve the signal to noise
ratio the height should be as minimal as possible. Its value
was chosen so as to satisfy the condition of the complete
absorption of alpha-particle with energy of 7.69 MeV leaving
the sample's surface in air or nitrogen at normal conditions.
In case when a sample under study is used as a collector there
would be background components in resulting spectra caused by
alpha-particles from the surface of drift volume, and by radon
decay and its daughter products in the working gas and on the
surface of the electrodes. To obtain starting low level of the
alpha-background proper, copper has been chosen as basic
construction material due to its low content of radioactive
elements. Total suppression of the background from the walls
could be achieved by putting the central part of the chamber
into anticoincidence mode with the circular part. The
background from the high-voltage electrode has been diminished
due to its grid structure having far less area than a solid
electrode. Moreover, absence of a partition wall between upper
and lower sections allows one to eliminate, in the
anticoincidence mode, alpha-particles generated in a gas and
having crossed the barrier of the high-voltage electrode, from
the spectrum. The upper section could be used to perform
independent measurement of radon content in the working gas.

Modeling has shown [4] that the additional selection of tracks
of the alpha-particles leaving the surface of the sample under
study could be performed using pulse shape analysis. An account
was taken of the variation of ionization density along track of
the alpha-particle, difference in drift velocities of positive
and negative ions, values of current pulses from both
components, depending on the distance of the track's to the
collector. Best predictions have been obtained with nitrogen as
a working gas due to its electronic conductivity in contrast to
air where negative charge is transported by molecular ions of
oxygen with captured electrons.

Maximum calculated drift times for positive and negative ions
in a drift gap for voltage of -2 kV and gas pressure of 620
torr  (mean atmospheric pressure at the level of BNO INR RAS)
are $1.57\cdot10^{-2}$ s; $1.26\cdot10^{-2}$ s and
$1.73\cdot10^{-2}$ s; $1.7\cdot10^{-5}$ s, for air and nitrogen
respectively. To register such long pulses there has been
developed a charge-sensitive low-noise preamplifier (CSP) with
optical feedback and self-discharge time of about $\sim 100$
ms. Signals from the four anodes IPIC are applied, through CSP
to the amplifiers with gain of 400. Then the signals are
divided into two ones; one signal goes to one of 16 inputs of
the analog-to-digital converter (ADC) NI6013 installed in the
personal computer (PC) and the other ones goes to one of the
inputs of the four-input low level discriminator (LLD). The LLD
generates output logic signal for the case where pulse's
amplitude at any input exceeds the preset threshold. A pulse
from LLD is applied to the input of the external triggering of
ADC. Sampling frequency for NI6013 was taken to be 2.5 kHz
(total sampling frequency is 10 kHz). Digitized pulses are
written down and stored in the PC.

\section{\label{sec3} Results of measurements}

\begin{figure}[hp]
\includegraphics[width=6.0cm,angle=270.]{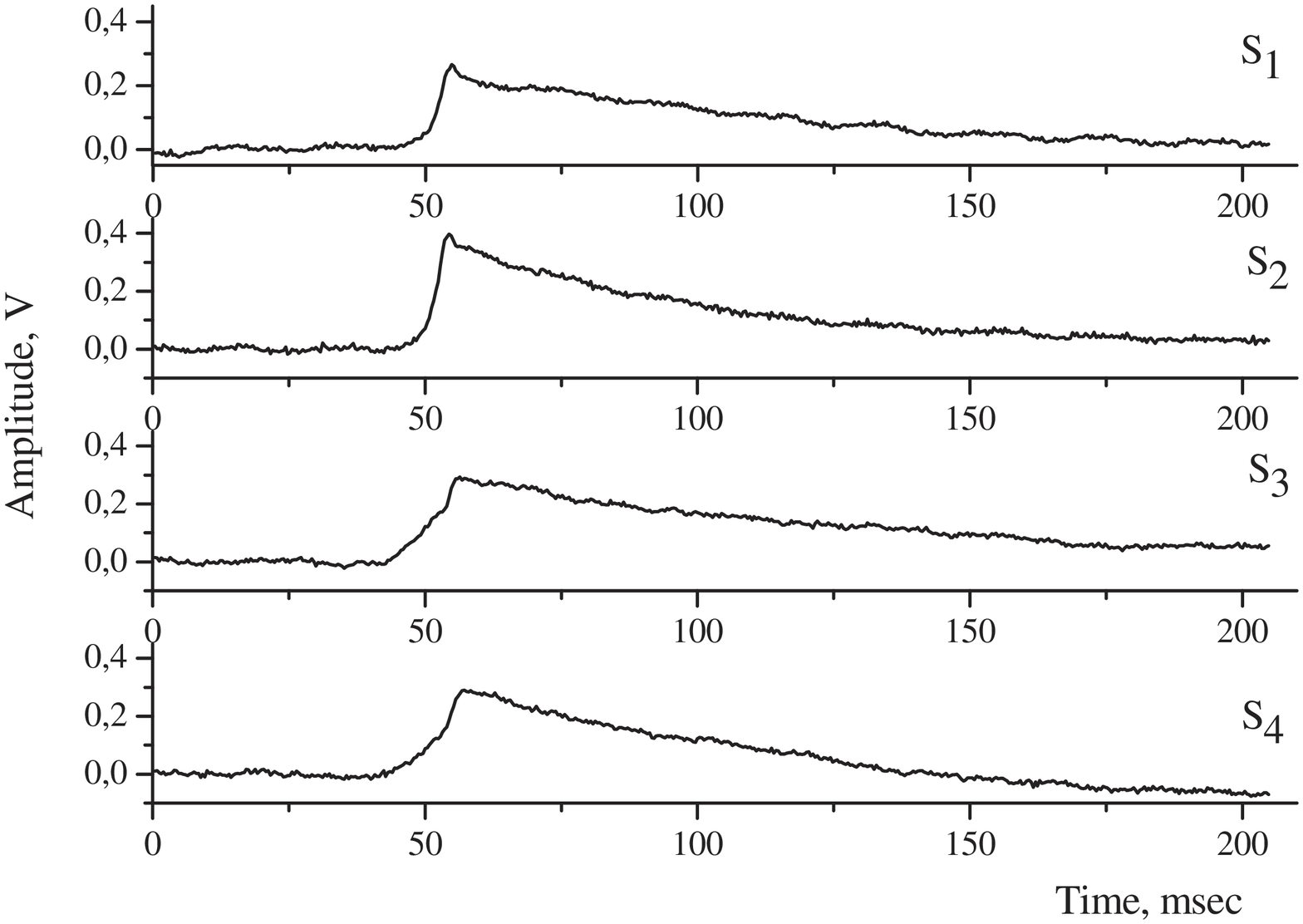}
\caption{Example of event where
alpha-particle has crossed all four sensitive areas (chamber is
filled with air).}
\label{pic2}
\end{figure}

Fig.\ref{pic2} demonstrates an event of an alpha-particle
crossing all four sensitive volumes. Numeration of plots
corresponds to that of anodes in fig.\ref{pic1}. The chamber
was filled with air.
\begin{figure}[hp]
\includegraphics[width=6.0cm,angle=270.]{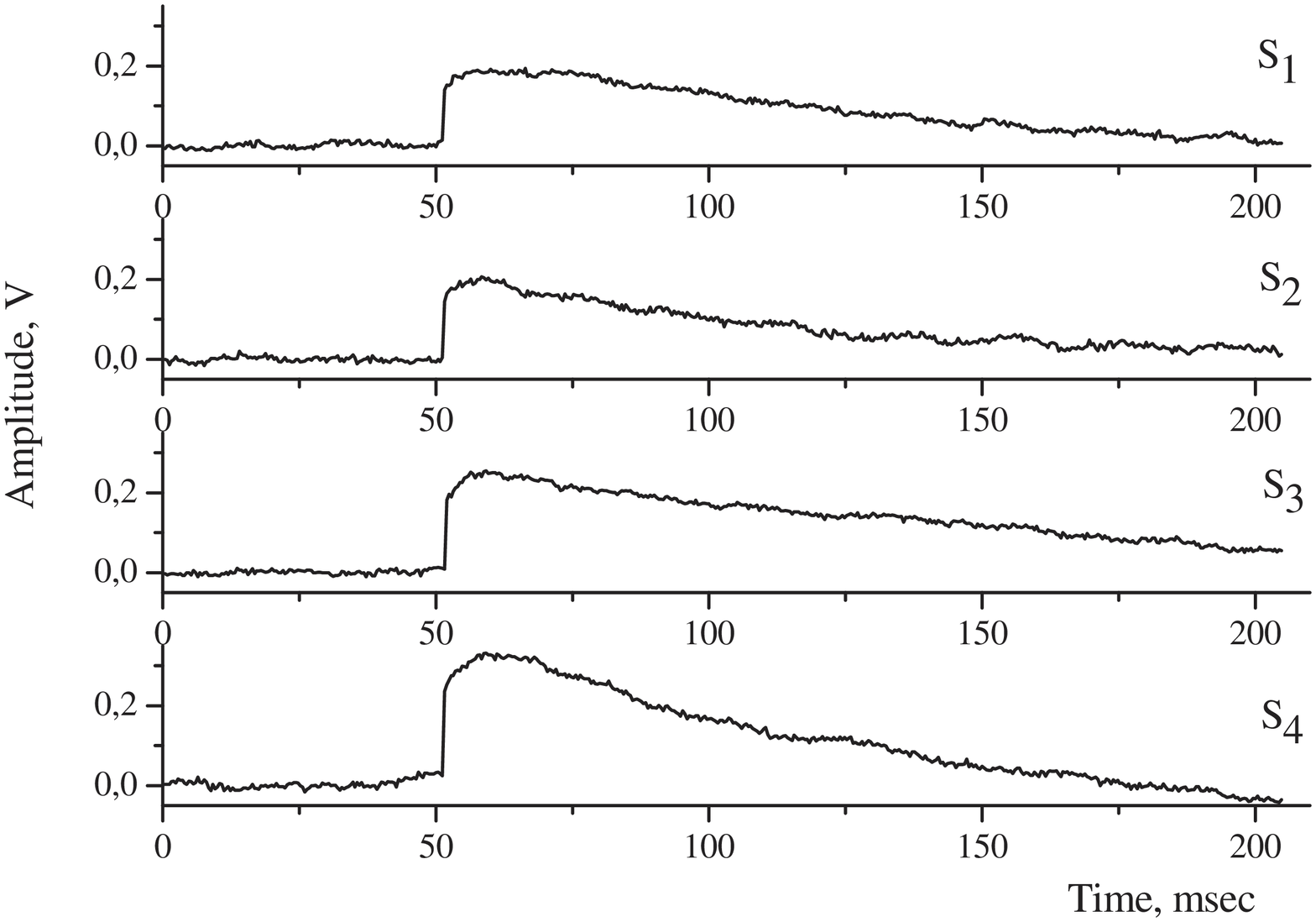}
\caption{Example of event where
alpha-particle has crossed all four sensitive areas (chamber is
filled with nitrogen).}
\label{pic3}
\end{figure}

Fig.\ref{pic3} illustrates the similar event for a case of IPIC
being blown through with liquid nitrogen vapour. Comparing
graphs of fig.\ref{pic2} and 3 one can see that the pulse rise
time in case of nitrogen is much shorter than for the air. This
distinction is related to the given above difference in drift
velocities of negative charge carriers. When nitrogen is not
properly purified from oxygen, both types of negative charge
carriers will take part in pulse formation, their ratio being
dependent on the distance of track's elements from the anode,
since the probability for electrically charged molecule of
oxygen to capture an electron is proportional to the path it
has traveled. To check the operating characteristics of IPIC an
$^{239}$Pu alpha-source (E$_\alpha \approx 5.157$ MeV) was
placed in the center of the cavity for samples. Its surface
(${\O}24$ mm) was covered with nickel foil 0.05 mm thick and a
hole of ${\O}1$ mm was made in the center of the foil which
provided alpha-particle count rate of $\sim1$ s$^{-1}$. The
chamber was blown with nitrogen during 3 hours ($^{~}30$ l).

\begin{figure}
\includegraphics[width=6.0cm,angle=270.]{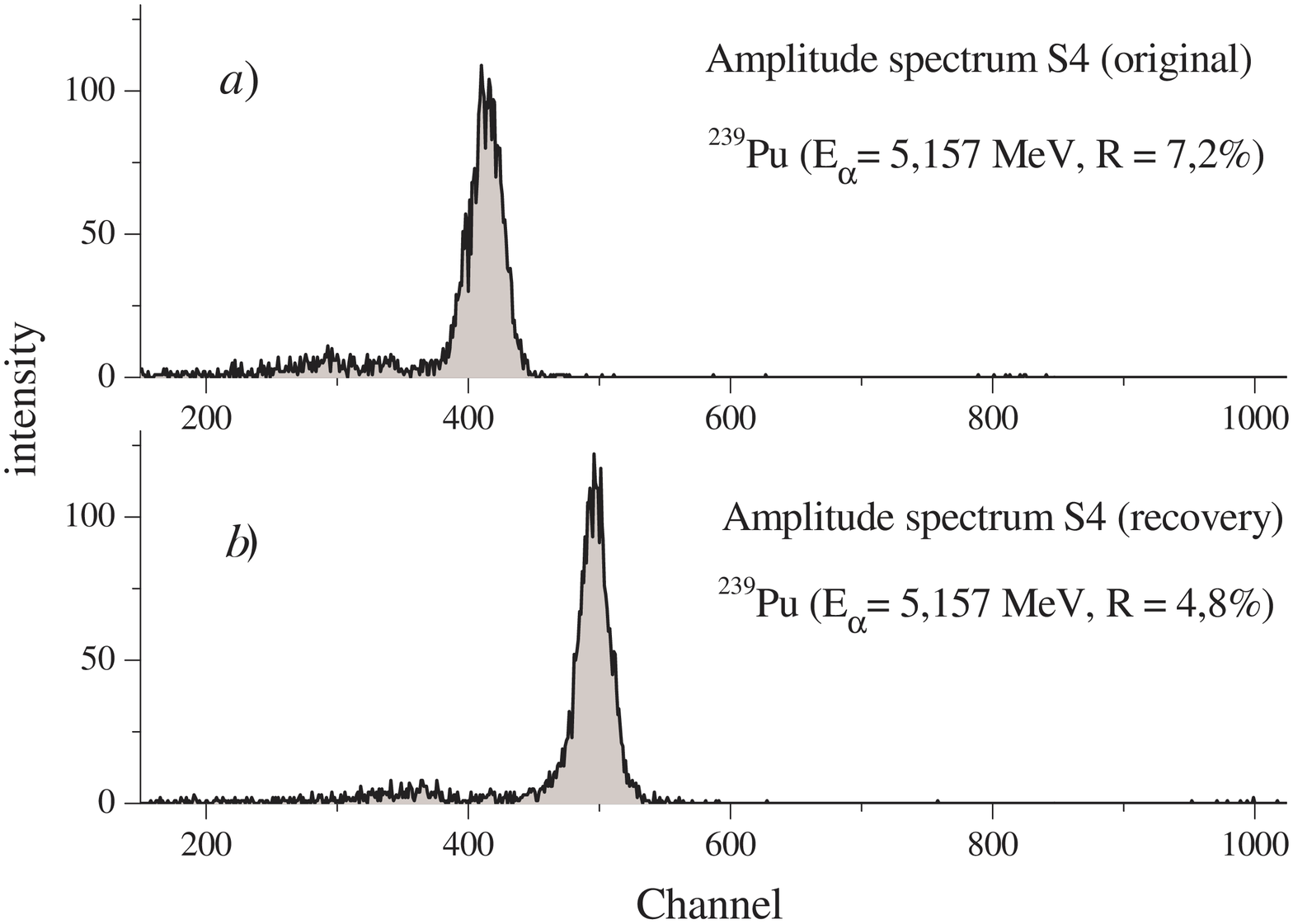}
\caption{$a$ - original spectrum of pulses taken from the lower central section (S4); 
$b$-recovered spectrum of pulses from the lower central section (S4).}
\label{pic4}
\end{figure}

Fig.\ref{pic4}$a$ shows a spectrum of pulses obtained during 70
min from the lower central section (S4). The spectrum is
plotted for charge pulse amplitudes at their maximum values.
Resolution is 7.2\%. The value of the amplitude at a pulse
maximum could differ for one and the same energy release since
this value for the finite CSP discharge decay time constant is
a function of current density. The true value of a charge
produced by an alpha-particle could be reconstructed by
introducing a correction for a CSP discharge through its
recalculation to the infinite decay time. With such a
correction the amplitude's value is determined by averaging it
over a given number of points taken from the time interval
following of the pulse current ending. Fig.\ref{pic4}$b$ shows
a spectrum of recovered pulses. The amplitude has been averaged
over ten points. Resolution has been improved up to 4.8\%. The
alpha-particle background for a sample of copper has been
carried out to investigate the IPIC measurement possibilities.
The sample was made in the form of a disc with 83.5 mm diameter
and 3 mm thickness. Its surface was treated with abrasive
materials and it was etched in nitrogen acid.

At first, the chamber was filled with atmospheric air
containing $^{222}$Rn at a level of  $\sim20$ Bq/m$^3$.
Total time to collect statistics made up 97 hours. A triggering
threshold was chosen to be 150 mV, which corresponds to the
alpha-particle energy of  $\sim1$ MeV. Total original
spectrum of amplitudes of not-reconstructed pulses from central
lower section (S4) is given in Fig.5a.
\begin{figure}
\includegraphics[width=6.0cm,angle=270.]{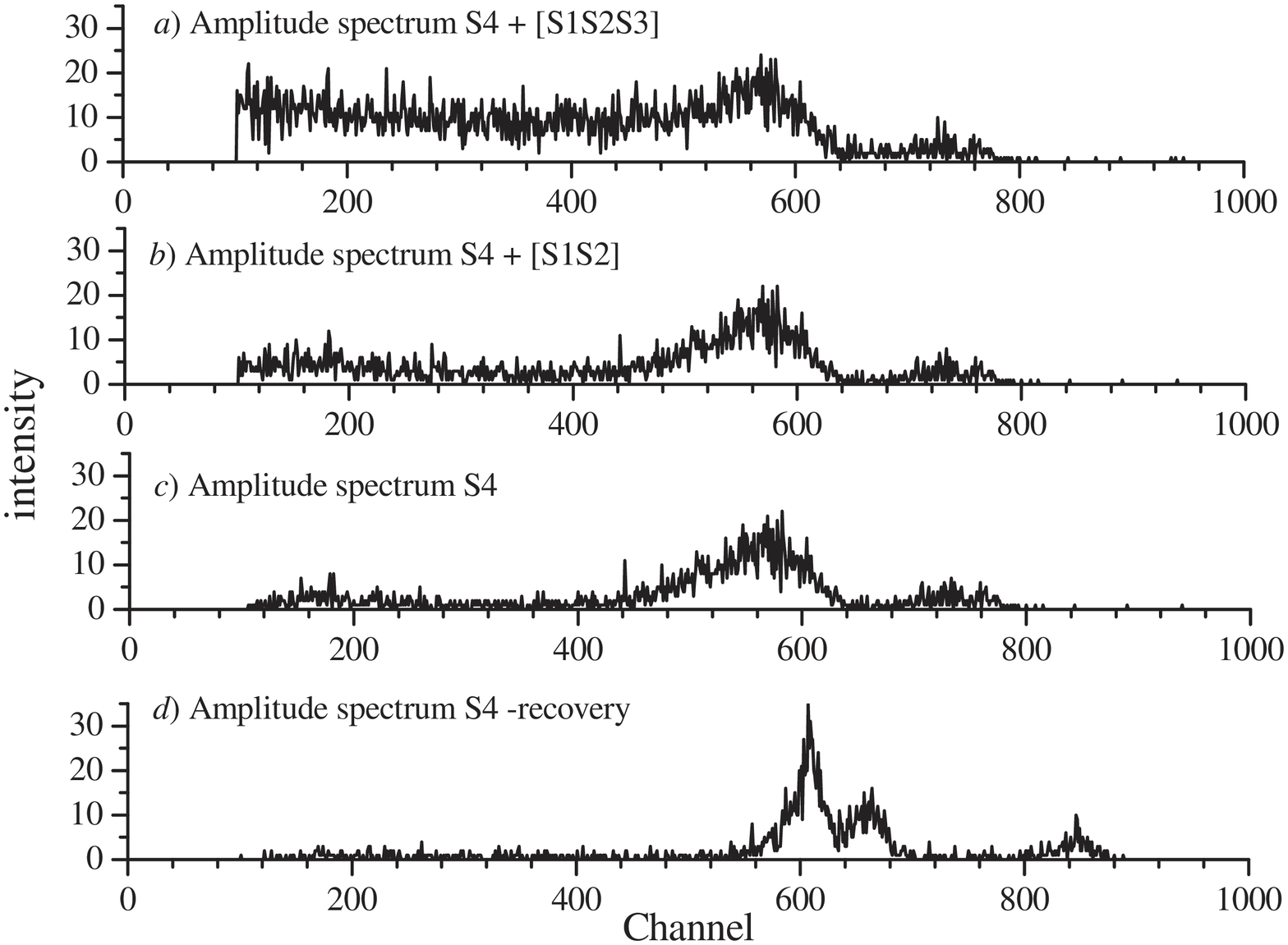}
\caption{Background spectrum of the
copper sample: $a$ - section S4 + sections S1S2S3; $b$ -
section S4 + sections S1S2; $c$ - spectrum of section S4; $d$ -
recovered spectrum of S4.}
\label{pic5}
\end{figure}
Spectrum in fig.\ref{pic5}$b$ illustrates the efficiency of
active shielding; the spectrum was plotted for pulses selected
from the spectrum of fig.\ref{pic5$a$} by excluding events
which have pulses with amplitudes higher than 0.5 MeV in the
lower circular section (S3). In fig.\ref{pic5}$c$, in addition
to this selection, another one has been added, that of exclude
events with pulses from the central upper (S1) and circular
(S2) sections. Fig.\ref{pic5}$d$ shows a spectrum of amplitudes
for the same pulses after introducing correction on the CSP's
self-discharge and after correction of microphonic noise caused
by high-voltage electrode vibrations by summing up noise from
S1 with pulse S4. Microphonic noises for these sections are in
antiphase.

The majority of pulses is due to alpha-decays of $^{222}$Rn
(T$_{1/2}$=3.82 days) in the air and its daughter $^{218}$Po
and $^{214}$Po on the high-voltage and collector electrodes S4.
Amplitudes of peaks of 6.00 MeV ($^{218}$Po) and 7.69 MeV
($^{214}$Po) correspond to alpha-particle energies. When
estimating the position of peak for 5.49 MeV ($^{222}$Rn), one
should take into account the recoil energy (0.10 MeV) which is
not entirely transmitted into ionization. In spectrum 5 d, the
position of peak from $^{222}$Rn decay corresponds to 5.54 MeV
using in its calculation the position of alpha-particle peaks
from $^{218}$Po and $^{214}$Po. Its resolution is 4.3\%. Ratio
of peak areas is determined by the efficiency of
alpha-particle's absorption in S4 and also by ratio of original
line intensities. $^{218}$Po and $^{214}$Po atoms are produced
with charge and they settle on the corresponding electrodes
under the electric field, so that half of alpha-paricles
generated at their decays goes to the wall and is absorbed
there. Ideally, ratio of peak areas for $^{222}$Rn , $^{218}$Po
and $^{214}$Po should be 2:1:1. In fig.\ref{pic5}$d$ it is
2:1.34:0.53.

\begin{figure}
\includegraphics[width=6.0cm,angle=270.]{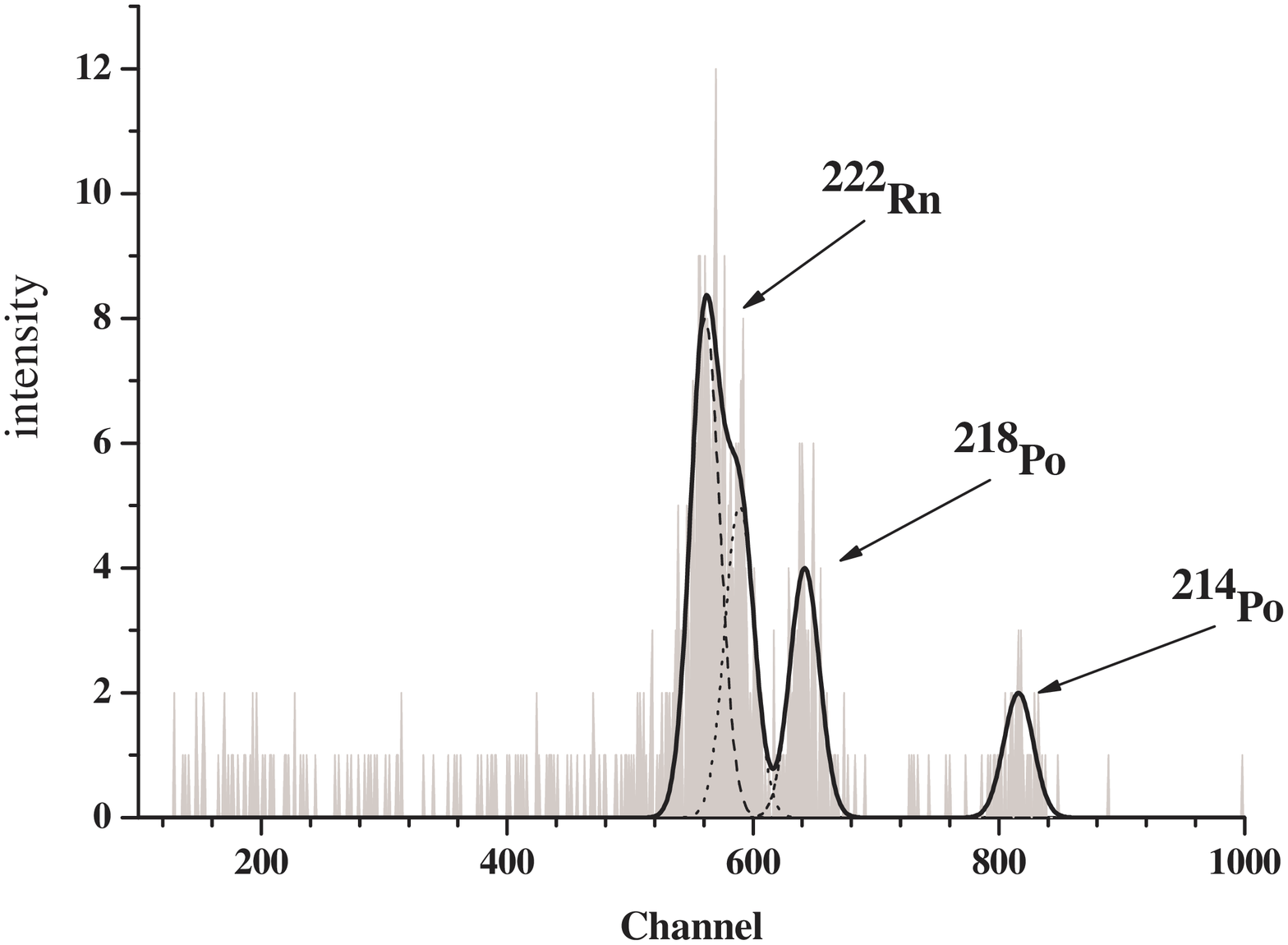}
\caption{Spectrum of pulse amplitudes
from Cu, collected during 316 hr and corrected for the
discharge and microphonic noise (blowing off with liquid
nitrogen vapour).}
\label{pic6}
\end{figure}

After measurements with air there were measurements carried out
with vapour of liquid nitrogen blowing through the IPIC ($\sim
10$ l/hr). Total time of collecting statistics was 316 hrs.
Fig.\ref{pic6} shows spectrum of pulse amplitudes from S4 with
correction for microphonic noise.

To analyze the spectrum let's start with peak $^{214}$Po
(channel 817, R=4,6\%). Its energy exceeds all the others and
is little subjected to distortions caused by admixture of other
lines, if to neglect the contribution of $^{212}$Po
(E$_\alpha=8.785$ MeV), the daughter product of decay of
$^{220}$Rn (T$_{1/2}$=55.6 s) of the series of $^{232}$Th, to
the alpha-particle background. The ratio of peak areas for
$^{222}$Rn, $^{218}$Po and $^{214}$Po should correspond to that
obtained from fig.\ref{pic5}$d$, in case when radon atoms decay
in the working gas. In such a case it is necessary to take into
account the difference in values of alpha-particle path in the
air and in nitrogen (the paths of an alpha-particle with energy
of 6 MeV and at pressure of 620 torr are 5.135 cm and 5.58 cm,
in air and nitrogen, respectively). Though, generally it is not
the case. Radon and its daughter products could be present in
the copper sample as a part of volume radioactive admixtures.
There is some microrelief on the surface of the sample; if the
characteristic size of its lugs is much less than the
alpha-particle path, then the energy loss due to these lugs
turns out to be small. Alpha-particles form the peak. The ratio
of areas of related peaks would differ from that obtained
above. To clarify the question concerning the type of related
peaks from $^{214}$Po source (internal or surface) one can use
data obtained with air. The events where coinciding pulses are
present only in sections S1 and S4 are produced, mainly, by the
alpha-particles that were generated in the gas due to radon
decay and crossed the border of high-voltage electrode. The
ratio of such events to the peak area of $^{214}$Po should be
constant when activity of radon changes, in case the latter is
uniformly distributed in a working gas. These ratios are
$1.57\pm0.15$ (for air) and $1.26\pm0.26$ (for nitrogen), which
coincide within the bounds of error. Therefore one can conclude
that the $^{214}$Po source is a surface one.

In fig.\ref{pic6}, one can see approximation of $^{222}$Rn ,
$^{218}$Po and $^{214}$Po peaks with Gaussians. The ratio of
their maximum positions corresponds to that in spectrum 5d. It
is clear that, in fig.\ref{pic6}, in addition to purely 'radon'
peaks there is also 5.297 MeV peak corresponding to
alpha-particles from $^{210}$Po decay (T$_{1/2}$=138.4 days).
It is produced in the chain reaction of $^{210}$Pb
(T$_{1/2}$=21.8 yrs) supposedly accumulated on the cathode
during adjustment and alignment while filling the IPIC with
ordinary air containing admixture of radon.

To determine surface alpha-activity (SAA) we use the range of
energy $1.1\div4.7$ MeV and assume that alpha-particle spectrum
in this range has a shape of flat step. Alpha-particles of  SAA
of the sample as well as a portion of peak alpha-particles that
has lost part of its energy in the material of the anode and
cathode give a contribution to this energy range. The relative
part of alpha-particles from radon decay and its daughter
products could be determined from fig.\ref{pic5}$d$. As it
seems impossible to evaluate the input of surface $^{210}$Po at
present stage of this experiment, the remaining background is
assumed to be due to SAA, which in such a case would be
($3.5\pm0.7$)$\cdot10^{-3}$ cm$^{-2}$hr$^{-1}$. The efficiency
of registration of alpha-particles with given energies was
taken to be a unit (disregarding edge effect). In order to find
a limit with 95\% C.L., this value of SAA was increased by 2
standard deviations.

When calculating the activity of a given isotope using its
measured alpha-activity, all the SAA was assumed to come from
this isotope.  Recalculation from the energy range under study
to the full range of all possible energy release is performed.
Upper limit of this range is determined as energy corresponding
to that of alpha-particle increased by 3 standard deviations.
Standard deviation value is determined by energy resolution.
The limit of energy range for $^{238}$U and $^{232}$Th was
determined as 5.16 MeV and 4.30 MeV, respectively.
Recalculation has been carried out under the assumption that
about 25\% of alpha particles that were generated in the layer
of the material with width equal to the length of this alpha
particle track, go out to the working gas [5].

The obtained results for activity and content of $^{238}$U and
$^{232}$Th in a copper sample calculated in the above described
way are given in Table 1.

SAA of light-reflecting film VM2000 which is supposed to be
used as a wall reflector in the cryostat with liquid argon in
the GERDA installation has been measured with IPIC [6].
Cryostat is used mainly to cool the assemblage of Ge-detectors
immersed into liquid. Liquid argon possesses scintillation
properties and can also be used as an anticoincidence-detector.
Wall reflector is necessary to improve conditions for light
collection.

\begin{figure}
\includegraphics[width=6.0cm,angle=270.]{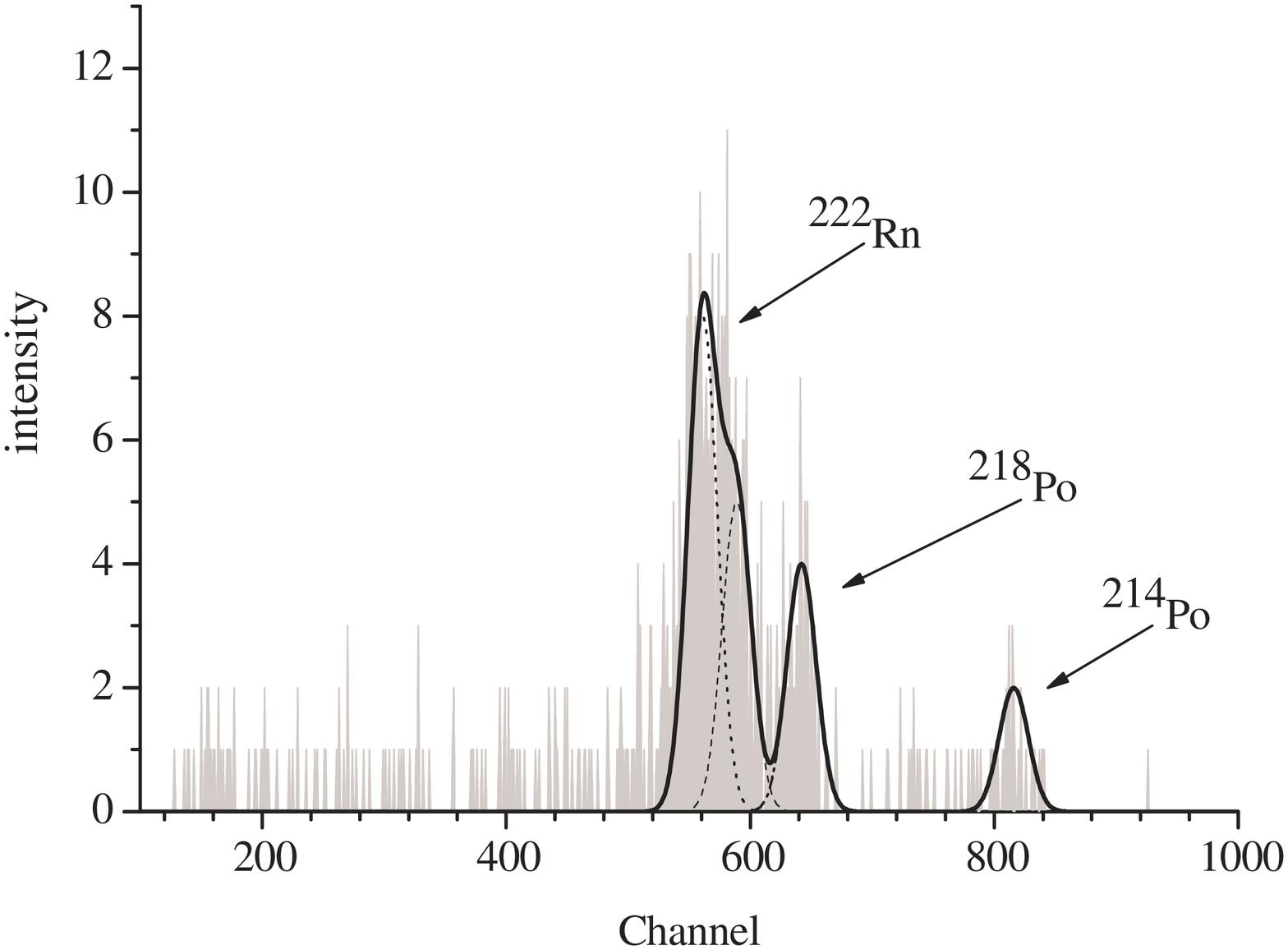}
\caption{Spectrum of pulse amplitudes
from VM2000, collected during 429 hr and corrected for the
discharge and microphonic noise (blowing off with liquid
nitrogen vapour).}
\label{pic7} 
\end{figure}

Sample of the film of 83 mm diameter and 70 mcm width was
placed at the bottom of the anode cavity in S4. The film being
of non-conductor material, it was covered with copper grid. In
the course of measurements IPIC was being blown with liquid
nitrogen vapors. Spectrum S4, collected during 429 hours is
given in fig.\ref{pic7}.
It is seen it is entirely similar to the spectrum in
fig.\ref{pic6}. After allowing for input of radon and its
daughter products the residual value of count rate in the range
of $1.1\div4.7$ MeV is ($3.0\pm0.6$)$\cdot10^{-3}$
cm$^{-2}$hr$^{-1}$, which coincide well with the data for the
copper sample within bounds of error. This could serve as an
indirect confirmation of the supposition that the residual
background is mainly due to $^{210}$Po presence on the cathode
grid. The contribution of proper SAA from the copper cathode
grid, alpha-active components being standard for the copper
sample, should be about 15 times less. This follows from the
ratio of surface areas of the grid and the sample.
\begin{figure*}[ht*]
\begin{center}
\includegraphics*[width=14.5cm,angle=0.]
{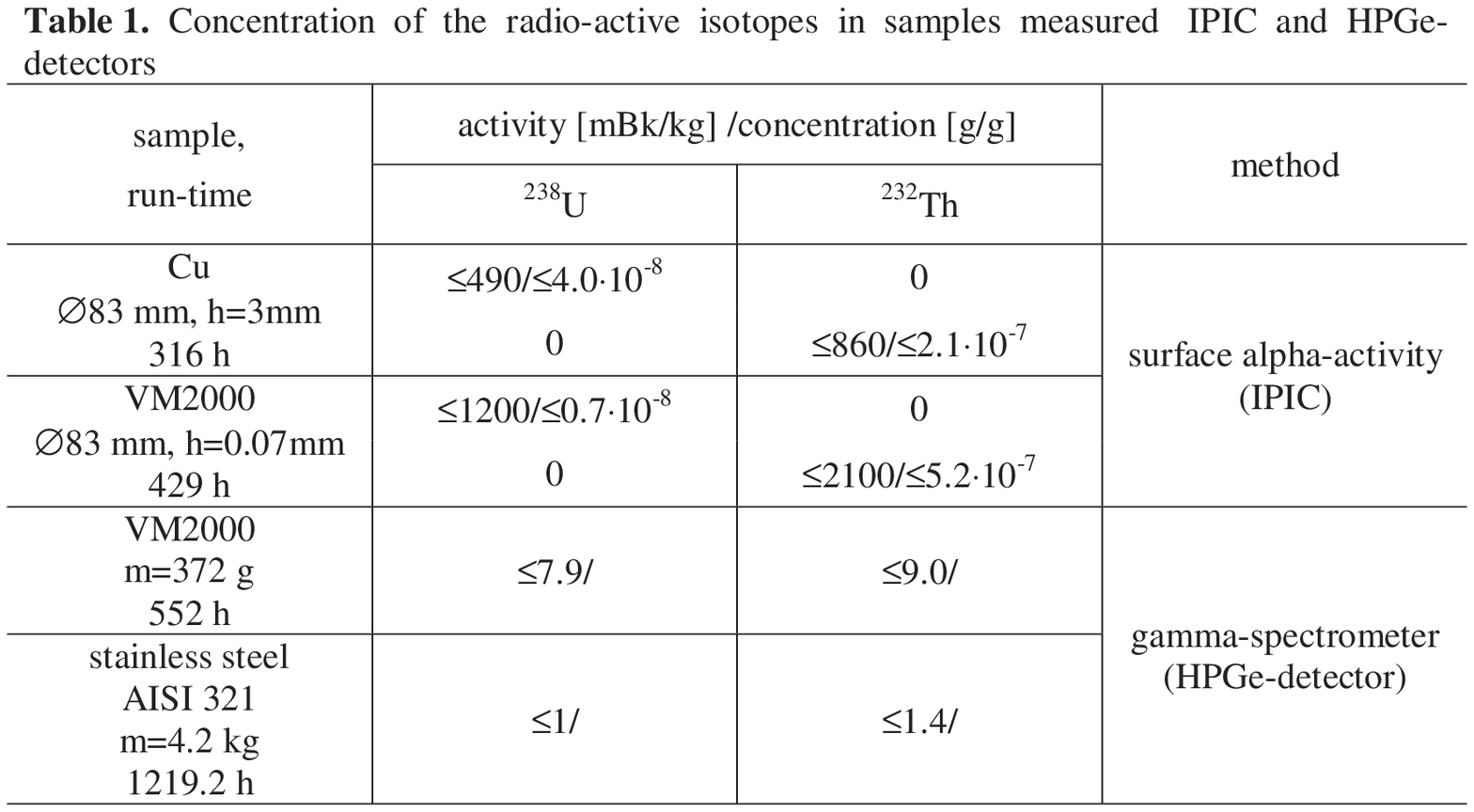}
\end{center}
\end{figure*}

The results of calculation for activities and content of
$^{238}$U and $^{232}$Th in the sample VM2000 are presented in
Table 1. Values for specific activity of $^{238}$U and
$^{232}$Th in samples of stainless steel  and VM2000 obtained
from the analysis of gamma spectra are given for comparison.
Measurements were carried out in the low background
installation 4HPGe with three semiconductor detectors made of
ultra-pure germanium [7]. The installation is located in the
underground low background laboratory of the BNO INR RAS at a
depth of 660 m w.e. The installation was used earlier in the
IGEX experiment to search for neutrinoless double beta-decay of
$^{76}$Ge.

The direct comparison of the results for VM2000 demonstrates in
a clear way that the sensitivity of IPIC achieved at this stage
is $\sim150$ times worse than that obtained with gamma method.
However, taking into account possible distortion of equilibrium
in the decay chains by 30 times or more would decrease the
difference down to $\leq5$ times.

The suggested above technique enables one to further increase
levels of sensitivity. For example, the background created by
$^{210}$Po in the cathode grid could be decreased by 100 or
more times by preliminary purification of the wires' surface
and taking measures to prevent its recurring contamination by
daughter products of the radon decay in the working gas. For
this purpose IPIC should be constantly re-filled with pure
nitrogen. The contribution of the SAA proper of the wires could
be significantly decreased by lessening the wire's diameter and
increasing spacing in the grid. For example, changing diameter
from $\O80$ mcm to  $\O20$ mcm and spacing from 2 mm to 4 mm
would decrease SAA of the cathode grid by 8 times. Background
of radon and its daughter products could be lowered by thousand
times, if  one would use nitrogen vapours preliminary passed
through the cooled charcoal trap to blow the IPIC. The pipes
for nitrogen vapour should be impermeable to radon of the air.
One should also remove from the fiducial volume of IPIC those
materials that have high emission of radon, e.g. elements of
the high-voltage divider.

Further enhancement of sensitivity could be achieved by
enlarging the area of the sample. At present, there are no
reasons preventing the increase of the IPIC's diameter by
$\sim3$ times. There is also an appealing possibility in
section S1 itself. In case the anode of this section is made
from guaranteed pure material (e.g., Si or Ge), the spectra of
this section could serve as a standard of zero background for
the analysis performed for section S4.

Moreover, to select tracks of alpha-particles starting out from
the sample's surface one could apply the analysis of the ion
component of the current pulse. A form of such a component
depends on the orientation of the track and distribution of
ionization density along the track. To make quantitative
analysis possible one need to improve noise performance of CSP
by about 3 times. Thereby energy resolution would be improved
as well.

Taking into account all the above mentioned factors the IPIC's
sensitivity could be risen by 100 times and even more. E.g.,
for alpha-particle count rate in the range of $1.1\div4.7$ MeV
being equal to 1/500 hr for the copper sample with surface of
400 cm$^2$ the sensitivity is up to 0.5 mBq/kg for $^{238}$U
and 0.9 mBq/kg for $^{232}$Th. Great advantage of this method
consists in the fact that in order to carry out measurements
one need not look for low background underground environment, 
the ordinary room would well accommodate this experiment. To
conclude authors thank the chief researcher Donchenko V.A. for
a fruitful interest in this work.

This work was fulfilled under the partial support of INTAS
program (grant INTAS-05-1-000008-7996 "The germanium detector
array for the search of neutrinoless double beta decay of
$^{76}$Ge at LNGS").


\begin{thebibliography}{12}

\bibitem{1} V.N. Gavrin, S.N. Danshin, A.V. Kopylov, V.I. Cherekhovsky
"Lowbackground semi-conductor gamma-spectrometer measuring
ultra-low concentration of $^{238}$U, $^{226}$Ra and
$^{232}$Th". Preprint INR RAS P-0494, Moscow, 1986. (in
Russian)

\bibitem{2} A.V. Kopylov, V.I. Cherekhovsky "Natural radioactivity of construction material"
Preprint INR RAS P-0604, Moscow, 1989. (in Russian)
\bibitem{3}V.V.Kuzminov. "Ion-pulse Ionization Chamber for Direct Measurement of a
Radon Concentration in the Air". Phys. of At. Nucl.,
V66,N3,2003, pp.462-465.

\bibitem{4}V.V. Kazalov (on behalf of the group)
"Identification of sources of alpha-particle background in Ion
Pulse Ionization Chamber", report presented at the Baksan Youth
School of Experimental and Theoretical Physics - 2005, 18-23
April 2005, Elbrus village, Kabardino-Balkaria, Russia. Edited
by KBSU, Nalchik, 2006, V.2, pp.101-108. (in Russian)

\bibitem{5} A.I. Abramov, Yu.A. Kazansky, Eu. S. Matusevich (Osnovy experimentalnyh metodov jadernoy fiziki)
"Basics of the experimental techniques in nuclear physics"
Atomizdat, Moscow, 1970, p.471 (in Russian)

\bibitem{6}GERDA - The GERmanium Detector Array for the search of neutrinoless $\beta\beta$ decays of $^{76}$Ge
 at LNGS. Proposal, INFN LNGS, 2004.

\bibitem{7}S.I.Vasiliev (for GERDA collaboration)
"Results of radiopurity measurements of the stainless steel samples AISI 321 and VM 2000 WLS
at the Baksan Neutrino Observatory (preliminary results)". The talk given on 26-28 June 2006
at Gran Sasso at the GERDA meeting.

\end{thebibliography}
\end{document}